\documentclass[twocolumn]{aastex63}

\usepackage{amssymb}
\usepackage{color}

\usepackage{listings}

\usepackage{xcolor}

\definecolor{codegreen}{rgb}{0,0.6,0}
\definecolor{codegray}{rgb}{0.5,0.5,0.5}
\definecolor{codepurple}{rgb}{0.58,0,0.82}
\definecolor{backcolour}{rgb}{0.95,0.95,0.92}

\lstdefinestyle{mystyle}{
    backgroundcolor=\color{backcolour},   
    commentstyle=\color{codegreen},
    keywordstyle=\color{magenta},
    numberstyle=\tiny\color{codegray},
    stringstyle=\color{codepurple},
    basicstyle=\ttfamily\footnotesize,
    breakatwhitespace=false,         
    breaklines=true,                 
    captionpos=b,                    
    keepspaces=true,                 
    numbers=left,                    
    numbersep=5pt,                  
    showspaces=false,                
    showstringspaces=false,
    showtabs=false,                  
    tabsize=2
}
\usepackage{hyperref}

\received{June 1, 2019}
\revised{January 10, 2019}
\accepted{\today}
\shorttitle{Multi-Scale Decomposition}
\shortauthors{Li}
\graphicspath{{./}{figures/}}

\begin{document}
\title{Multi-scale decomposition of astronomical maps -- a constrained diffusion method}

\correspondingauthor{Guang-Xing Li}
\email{ gxli@ynu.edu.cn}

\author{Guang-Xing Li}
\affiliation{South-Western Institute for Astronomy Research, Yunnan University,\\ Kunming, 650500 Yunnan, P.R. China}




\begin{abstract}
  We propose a new, efficient multi-scale method to decompose a map (or signal in general)
  into components maps that contain structures of different sizes. In the widely-used wave transform,  artifacts containing negative values arise around
  regions with sharp transitions due to the application of band-limited filters.  In our approach, the decomposition is achieved by solving a modified, non-linear version of the diffusion equation. This is inspired by the anisotropic diffusion methods, which establish the link between image filtering and partial differential equations. 
  In our case, the artifact issue is addressed where the positivity of
  the decomposed images is guaranteed. Our new method is particularly suitable for signals which contain localized, non-linear features, as typical of astronomical observations. It can be used to study the
  multi-scale structures of astronomical maps quantitatively and should be useful in observation-related
  tasks such as background removal. We thus propose a new measure called the ``scale spectrum'', which describes how the image values distribute among different components in the scale space, to describe maps. The method allows for input arrays of an arbitrary number of dimensions, and 
  a \texttt{python3} implementation of the
  algorithms is included in the Appendix and available at \url{https://github.com/gxli/constrained_diffusion_decomposition}. 
\end{abstract}

\keywords{ Wavelets  -- Multiscale transforms -- Image analysis 
}



\section{Introduction}
Many astrophysical processes are multi-scale in nature, and
many astronomical observations generate images that contain structures at a variety of scales. Being able to decompose such images/data into components of multiple
scales in a robust fashion would enable new approaches to data analysis and
interpretation. For example, such a decomposition would allow the study of the
relative importance of structures of different sizes. Multi-scale decomposition can also be useful for
observation-related tasks such as background removal. 

One commonly-used decomposition method is the wavelet
transform \citep[e.g.][]{cite-key}. In this method, the decomposition
is achieved by projecting an image onto orthonormal series generated by a
wavelet, where each projection would generate an image that contains structures
of a particular size.  Despite its success, in the wavelet transform,
the decomposed maps often contain artifacts in the vicinities of regions of
large contrasts, caused by the convolution with band-limited filters
\citep{Coifman95translation-invariantde-noising}. This is particularly problematic in astronomical problems, because the signals, such as the light from the sky or surface density distribution of astronomical objects, often exhibit significant, localized variations. Besides, in most astronomical or physical applications, the maps usually describe
the distribution of emission from the sky or the distribution of matter or
energy in space, and physically meaningful maps should contain positive values.
The maps generated by the wavelet transform usually contain negative values, which are consider unphysical in many cases. 

Methods like the multi-scale median transform have been
 developed and used \citep{1998ipda.book.....S,2011A&A...527A.145B}. Although
 some of these problems are addressed, the positivity of the results is not guaranteed. 
 \citet{2017MNRAS.464.4096L} proposed an iterative method where
 the positivity is guaranteed. However, the algorithm is difficult to implement
 and the decomposed maps contain some artificial discontinuities which are
 undesirable.  
 In this paper, inspired by a conceptual framework called the
 ``anisotropic diffusion" \citep{56205,Weickert96anisotropicdiffusion}, we propose a
 new method to perform such decompositions. 
 The method is robust and the
 positivity of the results is guaranteed, and  can be modified to achieve decompositions that suit different purposes.  

\begin{figure*}[htb!]
  \includegraphics[width=0.9 \textwidth]{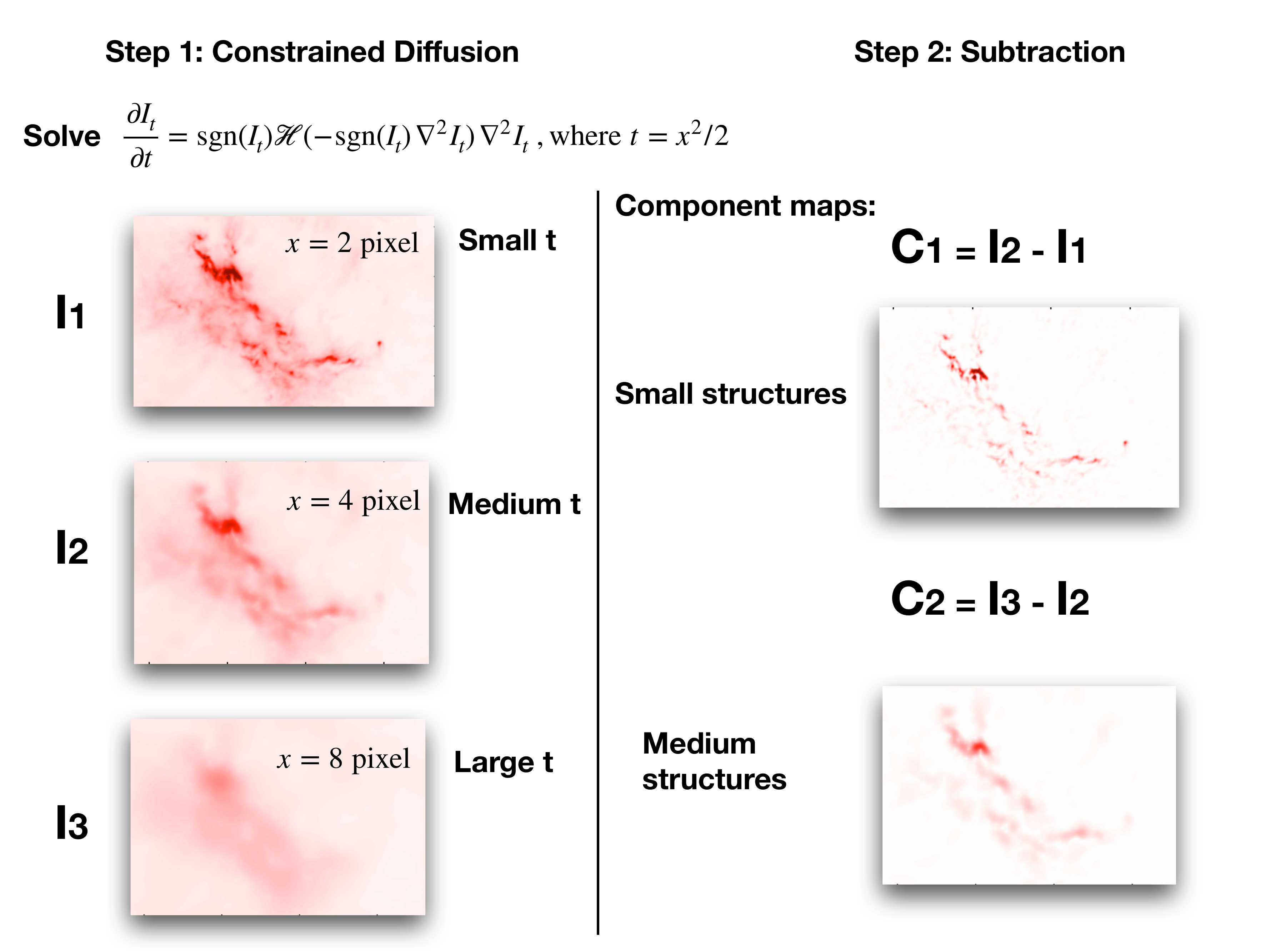} 
  \caption{An illustration of the overall procedure. See Sec. \ref{sec:method} for details.\label{fig:method} }
 \end{figure*}

 \section{Constrained Diffusion Method}\label{sec:method}
 In the wavelet transformation, the image is convolved with a set of Gaussian kernels of increasing size, and the decomposed maps are obtained by subtracting these smoothed images from one another. Similarly, in our method, given an input map $I(x, y)$, we produce
  a set of smoothed maps $I_l(x, y)$, and the decomposed maps will be constructed by subtracting the smoothed maps from one another. In the wavelet transform, the smoothing is achieved by convolving with a Gaussian kernel, and our major improvement is to perform this by solving Eq. \ref{eq:diffusion:v3}, which is a modified, non-linear version of the diffusion equation.
   A flowchart is presented in Fig. \ref{fig:method}. We note that in our method, the term  ``size'' corresponds to the dispersion of the Gaussian function.

\subsection{Diffusion-equation approach to image processing}
In image processing and computer vision, there exists a deep conceptual link between image filtering (such as smoothing) and partial differential equation. Convolving an image with a Gaussian kernel of dispersion $\sigma$ can be achieved by solving 
\begin{equation}\label{eq:pde}
  \frac{\partial I(x, y; t) }{\partial t} = \nabla^2 I(x, y; t)\;,
\end{equation}
where $I(x, y; 0) = I(x, y)$, and a proper smoothing will be achieved when $t=
l^2 / 2$ where $l$ is the smoothing length measured in terms of the dispersion\footnote{Recall that the solution to
$\partial  I/ \partial t =  \nabla^2 I$, $I(x, t=0) = \delta(x)$ is  Gaussian function whose dispersion $\sigma$ is related to $t$ by $t = \sigma^2 / 2$. }. 

Replacing Gaussian convolution with solving Eq. \ref{eq:pde} might seem to be an
unnecessary complication, as Gaussian convolution is much easier to implement
than solving a partial differential equation (PDE).  On a deeper level, the conceptual link
between image convolution/filtering and solving partial differential equations
has opened up the possibilities where one can develop new image operations by
providing the corresponding PDEs. In the past, this had led to the proposal of a
new group of techniques called anisotropic diffusion
\citep{56205,Weickert96anisotropicdiffusion}, where specially-designed PDEs are
used to achieve feature-preserving de-noising. In this paper, we import this
concept where we proposed to solve a modified version of the diffusion equation
to achieve multi-scale image decomposition.

\subsection{Constrained diffusion}

We propose to solve the following equation:
 \begin{equation} \label{eq:diffusion:v2}
  \frac{\partial I_t(x, y; t) }{\partial t} =  \mathcal{H}(-\nabla^2 I_t(x, y; t)) \nabla^2 I_t(x, y; t)\;, 
\end{equation}
where $t= l^2 /2 $ is ``effective time", and is related to the scale of interest $l$ by $t\sim l^2$, $\mathcal{H}$ is the Heaviside step function (the value of $\mathcal{H}$ is zero for negative arguments and one for positive arguments). 
The equation resembles diffusion equation, with the only difference lies in the introduction of the one additional Heaviside function.
It is the introduction of the Heaviside step function $\mathcal{H}(-\nabla^2 I_t(x, y; t))$ ensures that   $I_t(x, y; t)$  decreases monotonically with $t$, and this ensures the positivity of the decomposed maps. 

Eq. \ref{eq:diffusion:v2} works for input images which contain only positive values. To deal with images which also contain negative values,  we further modify Eq. \ref{eq:diffusion:v2} into 
 \begin{equation}  \label{eq:diffusion:v3}
  \frac{\partial I_t }{\partial t} ={\rm sgn}(I_t) \mathcal{H}({- \rm sgn}(I_t)  \nabla^2  I_t)   \nabla^2 I_t \;, 
\end{equation}
where $\rm sgn$ is the sign function (the mathematical function that extracts the sign of a real number). We insert two ${ \rm sgn}(I_t)$ functions to ensure that the results will be invariant under $I_t \rightarrow - I_t$. This guarantees that the regions with negative values will be processed in a self-consistent way. 

\subsection{Decomposed maps}
After obtaining the smoothed maps, the decomposed maps are derived by subtracting them against each other, e.g.  $C_{l \rightarrow l'} = I_{l}(x, y) - I_{l'}(x, y)$ would contain structures with sizes larger than $l$ yet smaller than $l'$.  As $I_{l}(x, y)$ decreases monotonically with increasing scale $l$, all of the decomposed maps $C$ are positive by construction.

\begin{figure*}
  
  \hspace*{0.6 cm}
  \includegraphics[width=0.95 \textwidth]{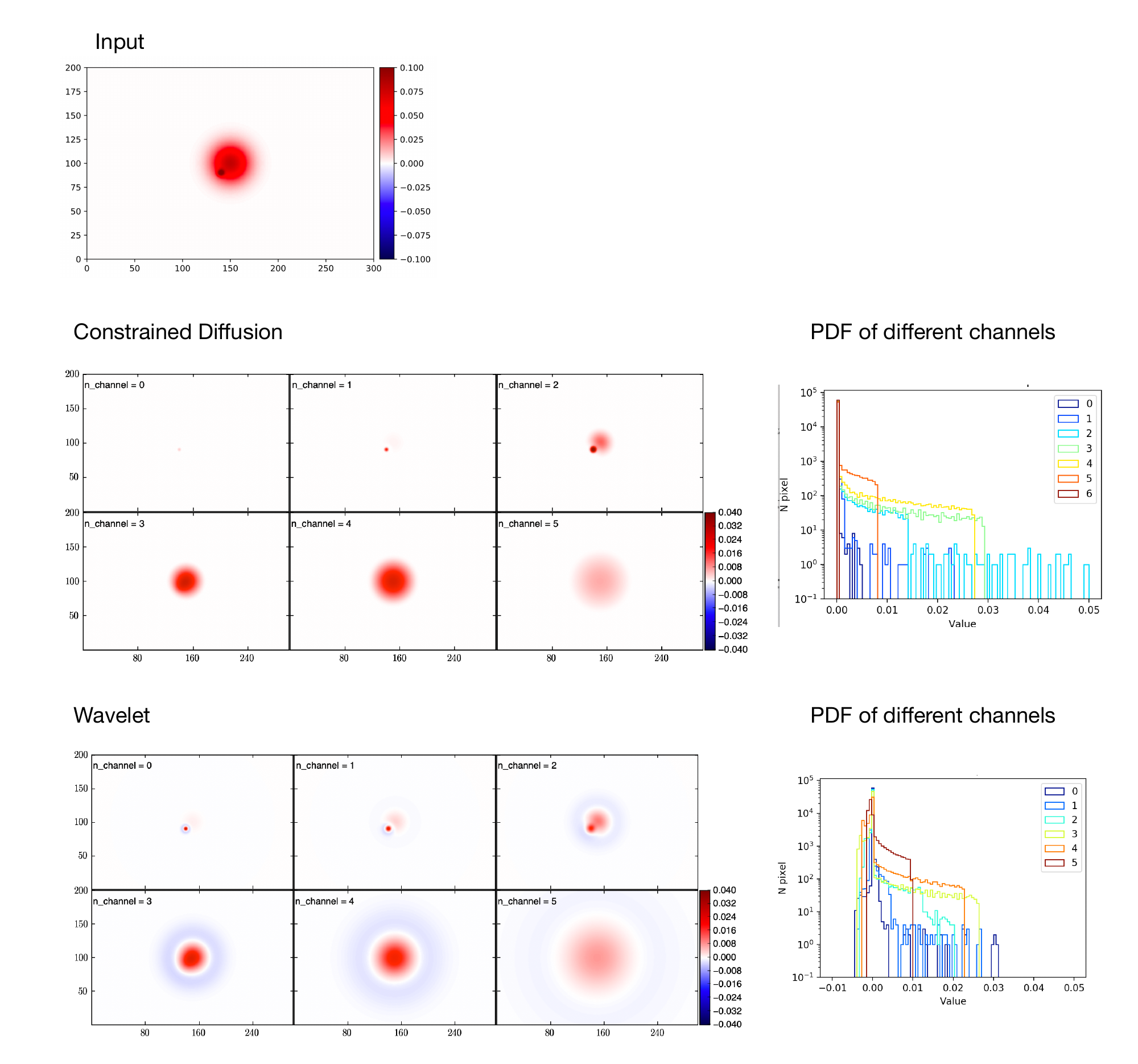} \\
  \caption{Comparison between the wavelet transform and our constrained diffusion. {\bf Top panel:}  Test data. It consists of two Gaussian functions. Their dispersions are 4 and 14 pixels, respectively. {\bf Middle panel:} results from the constrained diffusion. Channel $n$ contain structures of sizes  es are larger than $l = 2^{n}$ pixels  yet smaller than $l = 2^{n+1}$ pixels. {\bf Middle panel:} results from the wavelet transform.   \label{fig:comparison} } 
 \end{figure*}

 \section{Implementation \& Availability}

\subsection{Overview of Procedure}
 Our algorithm can be implemented for data with arbitrary number of dimensions.
 For simplicity, we consider the decomposition of a 2D map $I(x, y)$ of size $(n_x, n_y)$. 

\begin{enumerate}
  \item Defined a set of scales ranges upon which the image will be decomposed. For simplicity, we use $l = 2^n$ where 
  $n = 1, 2, 3, ...,  n_{\rm max}$ and $2^n_{\rm max}  \leq  {\rm min}(n_x, n_y) $.
  \item Taking $I(x, y)$ as the input, and evolve Eq. \ref{eq:diffusion:v3} from $t=1$ to $t =  l_{\rm max}^2 / 2 $, where $l_{\rm max} =  2^{n_{\rm max}}$. Register  the values of $I_n = I(x, y; t_n)$ where $t_n = 2^{2 n -1 } $. In the end, we have $I_0$, which is the original image, and $n_{\rm max}$ smoothed images.  
\item Generate the decomposed maps using $C_n = I_{n} - I_{n-1}$. $C_n$ represent structures whose sizes are larger than $l = 2^{n}$  yet smaller than $l = 2^{n+1}$. 
\end{enumerate}

After  the decomposition procedure, we obtain the  output, which is a set of images of the same size $(n_x, n_y)$. Note that by summing up the decomposed maps, the original input map can be recovered, e.g. $I(x, y) = \sum_n C_n(x, y)$. 

\subsection{Computation of smoothed images}
To generate the smoothed maps, we use a method we call ``constrained convolution". Our aim is to evolve Eq. \ref{eq:diffusion:v3} from $t=t_1$ to $t = t_2$, where we wish to achieve this in $n$ steps.  In each of our timestep, starting from $I_{t_1}$, we first compute the ``unconstrained" version of $I $ where
\begin{equation}
  I_{t_1 + \delta t}^{\rm unconstrained} = I_{t_1} \ast g(\sigma = \sqrt{2 \delta t})\;,
\end{equation}
where $g$ stands for the Gaussian function, and $\ast$ represents convolution. Based on this, the desired result can be achieved using
\begin{equation}
  I_{t_1 + \delta t} =   I_{t_1 + \delta t}^{+} +  I_{t_1 + \delta t}^{-} 
\end{equation}
where the positive part 
\begin{equation}
  I_{t_1 + \delta t}^{+}  = {\rm min}(  I_{t_1 + \delta t}^{\rm unconstrained} ,I_{t_1} ), {\rm for\, regions\, where}\, I_{t_1} > 0\;,
\end{equation}
and the negative part
\begin{equation}
  I_{t_1 + \delta t}^{-}  = {\rm max}(  I_{t_1 + \delta t}^{\rm unconstrained} ,I_{t_1} ), {\rm for\, regions\, where}\, I_{t_1} < 0\;
\end{equation}

Since our approach is effectively the Finite Difference theme where the truncation error is proportional to the step size \citep{1992nrfa.book.....P}, the  time step should be chosen such that $\delta_t / t < e_{\rm rel}$, where $ e_{\rm rel} << 1 $ is the desired accuracy.  
We adopt $ e_{\rm rel}  = 0.01 $. We further require that a physically meaningful timestep be smaller than yet comparable to unity, as our calculation is done on a rectangular grid and the accuracy of our calculations is limited by the discretization error. 
 Taking these into account, the minimum timestep is chosen to be 0.1. Because of this constraint, the accuracy of the first component map is limited to 0.1
A version of the code is provided in Appendix \ref{sec:code} where it has been tested using 1D, 2D, and 3D data. Due to fact that the results are invariant under $I \rightarrow -I$, regions containing negative values will be treated consistently. 

\section{Performance}


\subsection{Comparison to the wavelet transform}

To demonstrate the performance of our method, we first consider a simple situation where the input consists of two superimposed Gaussian functions. These Gaussian functions have dispersions of 4 and 14 pixels, respectively. 
Fig. \ref{fig:comparison} presented the results from wavelet decomposition
and our constrained diffusion method. Details concerning the implementation of the wavelet transform can be found in Appendix \ref{sec:appendix:wavelet}.
From the wavelet decomposition results,
one can identify the presence of artifacts around regions of sharp transitions (e.g. regions surrounding the narrower Gaussian) caused by the application of band-limited filtering. Besides,  although the input is everywhere positive, after performing the wavelet decomposition, the decomposed maps contain coherent regions of made of negative values. 

In Fig. \ref{fig:comparison} we present the decomposition achieved with our constrained
diffusion method. Compared to the wavelet transform, the different Gaussian
components have been separated into different channels representing structures of different sizes. Different from the wavelet transform, the contained diffusion
decomposition provides a clean separation, e.g. these Gaussian functions can be separated in a way that they do not
interfere with each other and the decomposed maps do not contain negative values. Moreover,  The channel within which the structure is contained can thus be directly related to its size. 
 E.g. the $\sigma = 4\rm \; pixel$ component is most prominent in the $n=2$ channel, which, according to our definition, contain structures of sizes $4 <l < 8$ pixels, and the $\sigma = 14\rm \; pixel$ component is most prominent in $n=3$ channel, which contains structures of sizes $8 <l < 16$ pixels.

\subsection{Applications to 1, 2 and 3-dimensional data}








 \begin{figure*}[htb!]
  \includegraphics[width=1\textwidth]{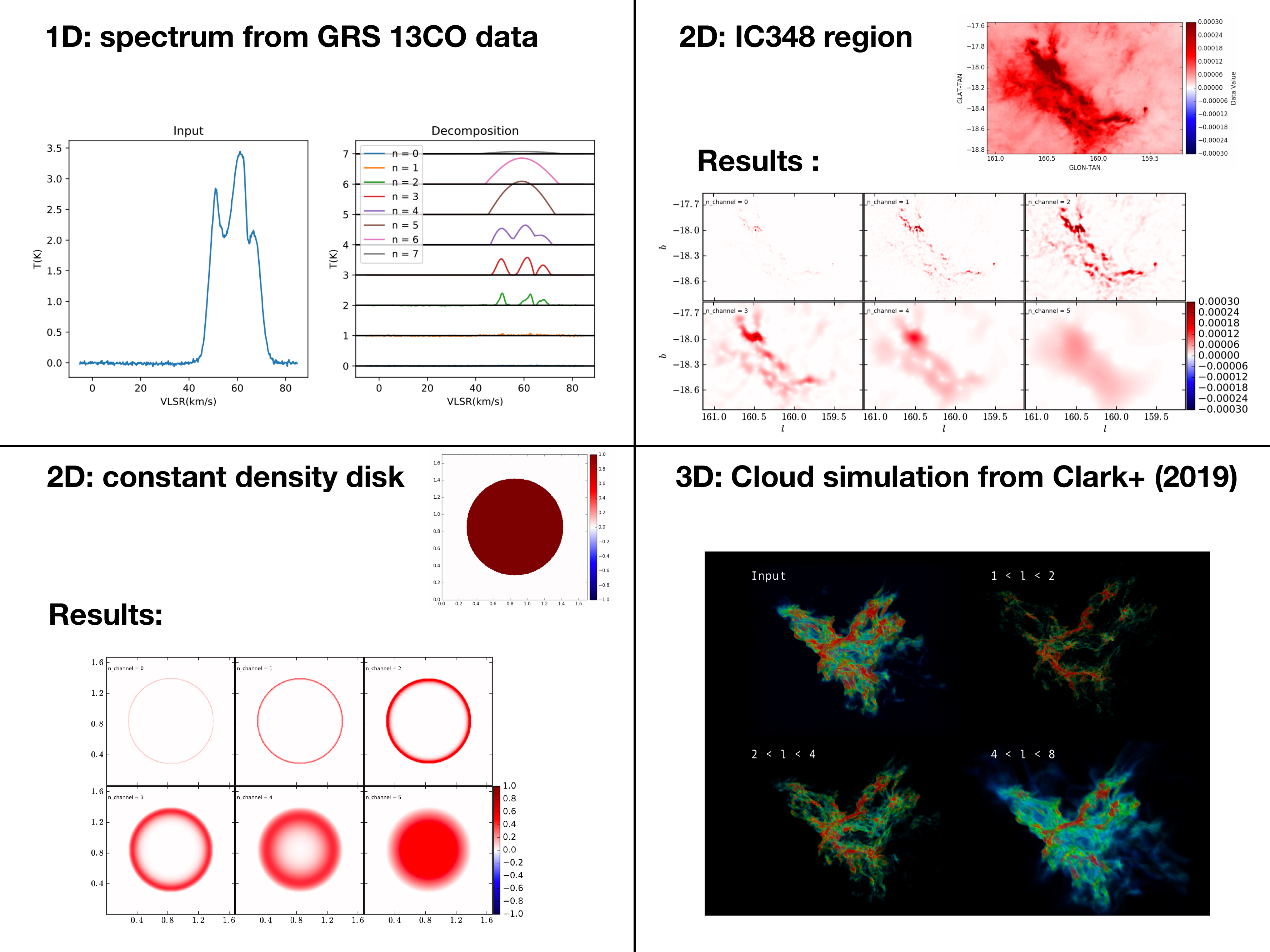} 

  \caption{Sample example applications of our constrained diffusion decomposition. In the 1D case, we perform a decomposition to a $^{13}$CO spectrum from the GRS survey \citep{2006ApJS..163..145J}. In the 2D case, we perform decomposition to a constant density 2D disk, and to a surface density map of the IC348 region using data from \citep{2016A&A...587A.106Z}. In the 3D case, we perform decomposition to a 3D density structure of a molecular cloud taken from a simulation performed by \citet{2019MNRAS.486.4622C}. Enlarged version of some of these figures as well as comparisons with wavelet decompositions can be found in  Appendix \ref{sec:plots}.  \label{fig:123}}
 \end{figure*} 
 Our algorithm is general and can be applied to an arbitrary number of dimensions. Some examples are presented in Fig. \ref{fig:123}.

 In the 1D case, we extracted a $^{13}$CO spectrum around  the central region of W43
 using data from the Galactic Ring Survey data \citep{2006ApJS..163..145J}, and performed a decomposition. In the results, the
 high-frequency noise, two narrower components, and a broad component can be
 separated into different channels. In the 2D case, we consider a first example where we perform
 decomposition to a 2D disk of a constant surface density. In the results,  the edge of the disk is contained in the first channels of
 smaller $l$, reflecting the presence of the sharp
 transition at its boundary, and the middle part of the disk is separated into
 channels of larger $l$. We also perform a decomposition on
  the surface density distribution of the IC348 region using data from 
 \citet{2016A&A...587A.106Z}. Our decomposition can
 separate the observed structure into different components. In the 3D case, we constructed a 3D density cube
 using simulation data from  \citep{2019MNRAS.486.4622C}. The simulations have produced a
 molecular cloud with a highly complex density structure, and the decomposition allows us
 to separate it into different components.

\section{Scale spectrum -- Quantifying multi-scale structures}

\begin{figure*}[htb!]
  \includegraphics[width=1\textwidth]{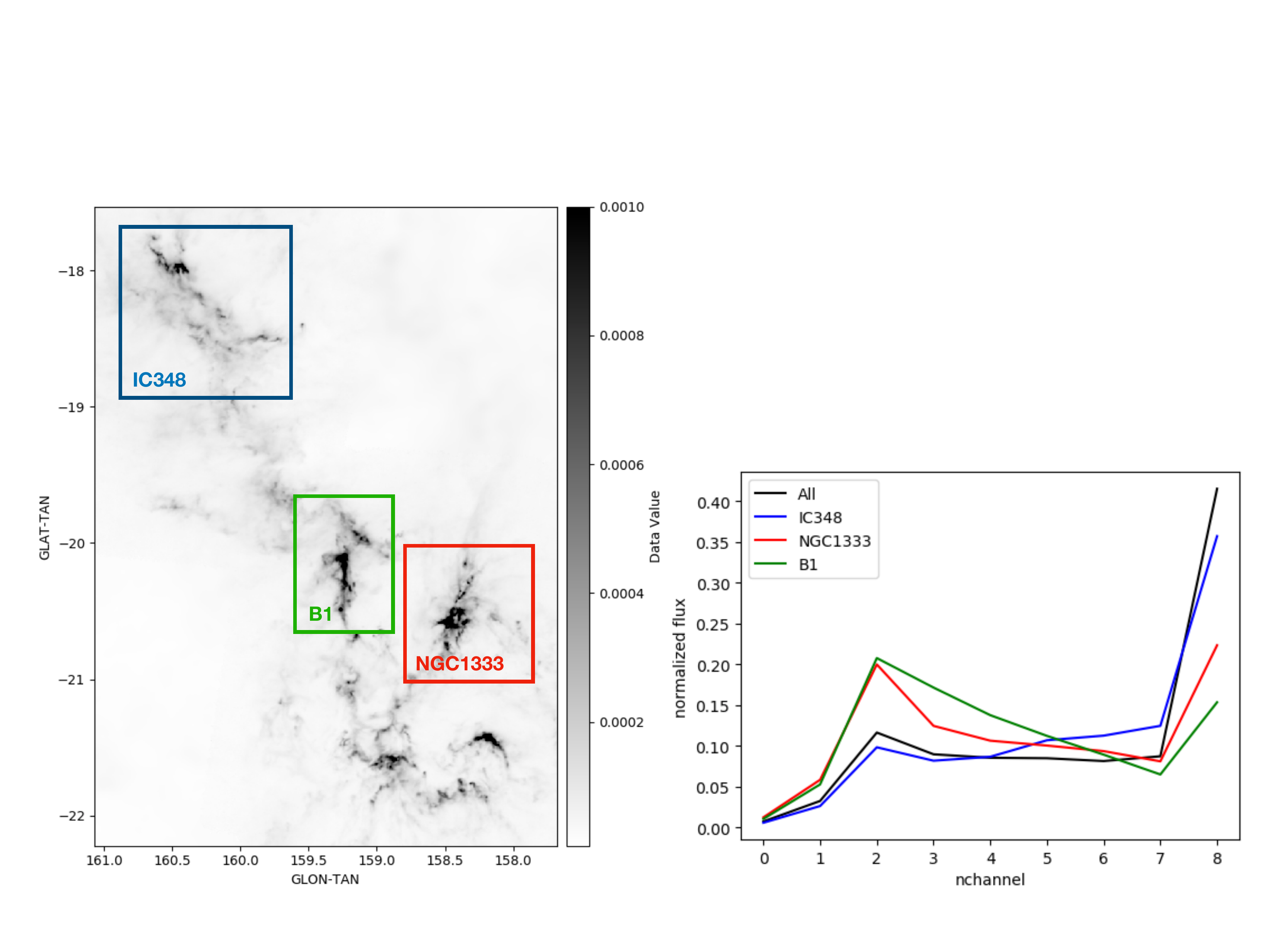} 

  \caption{Scale spectrum of regions in the Perseus molecular cloud. {\bf Left Panel:} surface density distribution of the Perseus molecular cloud constructed using data taken from \citet{2016A&A...587A.106Z}. The boundaries of some regions are indicated.  {\bf Right Panel:} Scale spectrum of Perseus cloud and the sub-regions. The $x$-axis is the channel number, where 
  channel $n$ contain structures of sizes  es are larger than $l = 2^{n}$ pixels  yet smaller than $l = 2^{n+1}$ pixels. The very last channel in the plot contains the residuals, e.g. all structures that are too large to be contained in the second last channel. The $y$-axis is the normalized mass. \label{fig:scale:spectrun}
  }
 \end{figure*} 
 The evolution of molecular clouds is a multi-scale process. As a result, the density structures of molecular clouds are made of substructures that distribute over a continuous range of scales. Quantifying these multi-scale density structures would provide insights into the underlying physics. Based on our decomposition, we propose a measure called ``scale spectrum'' to achieve this. For a given region $R$,  the scale spectrum is defined as
 \begin{equation}
   I_l =  \int_R I_l(x, y)\; {\rm d} s \;, 
 \end{equation}
 which is a representation of how matter distributes in structures of
 different scales. In a smooth region, matter is distributed into structures of
 large $l$, whereas in a structured region, matter distributes in structures
 of large $l$. Fig \ref{fig:scale:spectrun} plots the scale spectra of different
 subregions in the Perseus molecular cloud, where gas in e.g., the NGC1333 region
 tend to distribute in channels with smaller $l$ in comparison to other e.g. the
 IC348 region. This agrees with our current understandings, as  NGC1333 is
 the most evolved region in the cloud with active star formation where we expect more structures
 \citep{2008hsf1.book..346W,2005A&A...440..151H}. The scale spectrum offers a
 quantitative way to describe a region, which enables quantitative comparisons.




\section{Conclusion  Remarks}
We propose a new method to decompose an image/signal into a set of maps
that contain structures at different scales in a robust way. The method
consists of producing smoothed images using a modified, non-linear version of
the diffusion equation, and the decomposed images are obtained by subtracting
these smoothed images from one other. Compared to the widely-used wavelet
transform, our method is robust against the presence of sharp transitions. If
the input image is positive, the positivity of the decomposed maps is
guaranteed. The positivity is often necessary or desirable for
subsequent analyzes and interpretations. The method is implemented as a
\texttt{python3} program  available in Appendix \ref{sec:code}. A maintained
version is available through
\url{https://github.com/gxli/constrained_diffusion_decomposition}. The method is
applicable for n-dimensional data. We
expect our decomposition to be useful in analyzing and quantifying images, and inpractical tasks such as data
visualization and background removal.

This algorithm allows us to propose a new measure called the ``scale spectrum'' to quantify images. 
It describes how the observed values distribute in substructures of
different sizes, leading to a a new, quantitative way to describe and compare
structures. 

This paper is also an introduction of the anisotropic diffusion method to the 
astronomical community. In this new approach, decomposition is achieved by
solving a modified diffusion equation. This conceptual link should enable other new, meaningful
ways to decompose maps by providing similarly-modified partial differential equations to achieve different purposes. 

  \section*{Acknowledgements}
Guang-Xing Li acknowledge supports from NSFC grant W820301904 and 12033005.

  \appendix
\section{Implementation and Usage}\label{sec:code} A python3 implementation of
our method can be found below (constrained\_diffusion\_decomposition.py).
\begin{lstlisting}[language=Python]
#!/usr/bin/env python
from astropy.io import fits
import sys
import numpy as np
from math import log
from scipy import ndimage


def constrained_diffusion_decomposition(data,
                                      e_rel=3e-2,
                                      max_n=None, sm_mode='reflect'):

    """
        perform constrained diffusion decomposition
        inputs:
            data: 
                n-dimensional array
            e_rel:
                relative error, a smaller e_rel means a better
                accuracy yet a larger computational cost
            max_n: 
                maximum number of channels. Channel number
                ranges from 0 to max_n
                if None, the program will calculate it automatically
            sm_mode: 
                {'reflect', 'constant', 'nearest', 'mirror', 'wrap'}, optional The mode
                parameter determines how the input array is extended beyond its
                boundaries in the convolution operation. Default is 'reflect'.
        output:
            results: of constained diffusion decomposition. Assuming that the input
            is a n-dimensional array, then the output would be a n+1 dimensional
            array. The added dimension is the scale. Component maps can be accessed
            via output[n], where n is the channel number.

                output[i] contains structures of sizes larger than 2**i pixels
                yet smaller than 2**(i+1) pixels.
            residual: structures too large to be contained in the results
                
    """
    
    ntot = int(log(min(data.shape))/log(2) - 1)    
    # the total number of scale map
    
    result = []
    # residual = []
    # residual.append(data)
    if  max_n is not None:
        ntot = np.min(ntot, max_n)        
    print("ntot", ntot)

    diff_image = data.copy() * 0

    for i in range(ntot):
        print("i =", i)
        channel_image = data.copy() * 0  

        # computing the step size
        scale_end = float(pow(2, i + 1))
        scale_begining = float(pow(2, i))
        t_end = scale_end**2  / 2  # t at the end of this scale
        t_beginning = scale_begining**2  / 2 # t at the beginning of this scale

        if i == 0:
            delta_t_max = t_beginning * 0.1
        else:
            delta_t_max = t_beginning * e_rel



        niter = int((t_end - t_beginning) / delta_t_max + 0.5)
        delta_t = (t_end - t_beginning) / niter
        kernel_size = np.sqrt(2 * delta_t)    # size of gaussian kernel
        print("kernel_size", kernel_size)
        for kk in range(niter):
            smooth_image = ndimage.gaussian_filter(data, kernel_size,
                                                   mode=sm_mode)
            sm_image_1 = np.minimum(data, smooth_image)
            sm_image_2 = np.maximum(data, smooth_image)

            diff_image_1 = data - sm_image_1
            diff_image_2 = data - sm_image_2

            diff_image = diff_image * 0

            positions_1 = np.where(np.logical_and(diff_image_1 > 0, data > 0))
            positions_2 = np.where(np.logical_and(diff_image_2 < 0, data < 0))

            diff_image[positions_1] = diff_image_1[positions_1]
            diff_image[positions_2] = diff_image_2[positions_2]

            channel_image = channel_image + diff_image

            data = data - diff_image
            # data = ndimage.gaussian_filter(data, kernel_size)     # !!!!
        result.append(channel_image)
        # residual.append(data)
    residual = data
    return result, residual


if __name__ == "__main__":
    fname = sys.argv[1] 
    hdulist = fits.open(fname)
    data = hdulist[0].data
    data[np.isnan(data)] = 0
    result, residual = constrained_diffusion_decomposition(data)

    nhdulist = fits.PrimaryHDU(result)
    nhdulist.header = hdulist[0].header
    nhdulist.header['DMIN'] = np.nanmin(result)
    nhdulist.header['DMAX'] = np.nanmin(result)

    nhdulist.writeto(sys.argv[1] + '_scale.fits', overwrite=True)
\end{lstlisting} and a maintained version is available at \url{https://gxli.github.io/Constrained-Diffusion-Decomposition/}.
The code can be used as a python package
\begin{lstlisting}[language=Python]
# assuming that data is a N-dimensional array 
import  constrained_diffusion_decomposition as cdd
# the decomposition can be computed via
result, residuals = cdd.constrained_diffusion_decomposition(data)
\end{lstlisting}

and as a \texttt{shell} script

\begin{lstlisting}[language=sh]
# assuming the data is contained in the input.fits
python3  constrained_diffusion_decomposition.py input.fits 
# the output will be written as input.fits_scale.fits
\end{lstlisting}
Given a input $I$ of size $(n_x, n_y)$, the output $C$ takes the form of  of  $(n_l, n_x, n_y)$ where decomposed maps are arranged such that $C[i]$ has same size as the input, and it contain structures of sizes than range from  $2^i$ to $2^{i +1}$ piexels. The ``size'' corresponds to the dispersion of the a Gaussian function. The \texttt{residual} contains the residuals e.g. structures whose sizes are even larger than $2^{i +1}$ pixels. 
  The program 
has been tested with 1, 2 and 3-dimensional data.

\section{The wavelet transform}\label{sec:appendix:wavelet}
We compare our method against a version of the wavelet transform called the {à} Trous wavelet transform \citep{starck2008astronomical}. 
The wavelet decompositions is a procedure where the decompositions is achieved by projecting an image onto a set of Hilbert basis, where different basis correspond to structures at different scales. In a simple case, the wavelet transform can be realized by a set of convolutions where the wavelet component $n$ is define as 
\begin{eqnarray}\label{eq:wavelet}
  I_n(x, y) = I(x, y) \ast (g(2^n) - g(2^{(n-1)}))\;,
\end{eqnarray}
where $I(x, y)$ represents the original emission map and $g(x)$ represents the Gaussian function where $x$ is the dispersion.

\section{Performance demonstration}\label{sec:plots}
\begin{figure}
  \includegraphics[width=0.9 \textwidth]{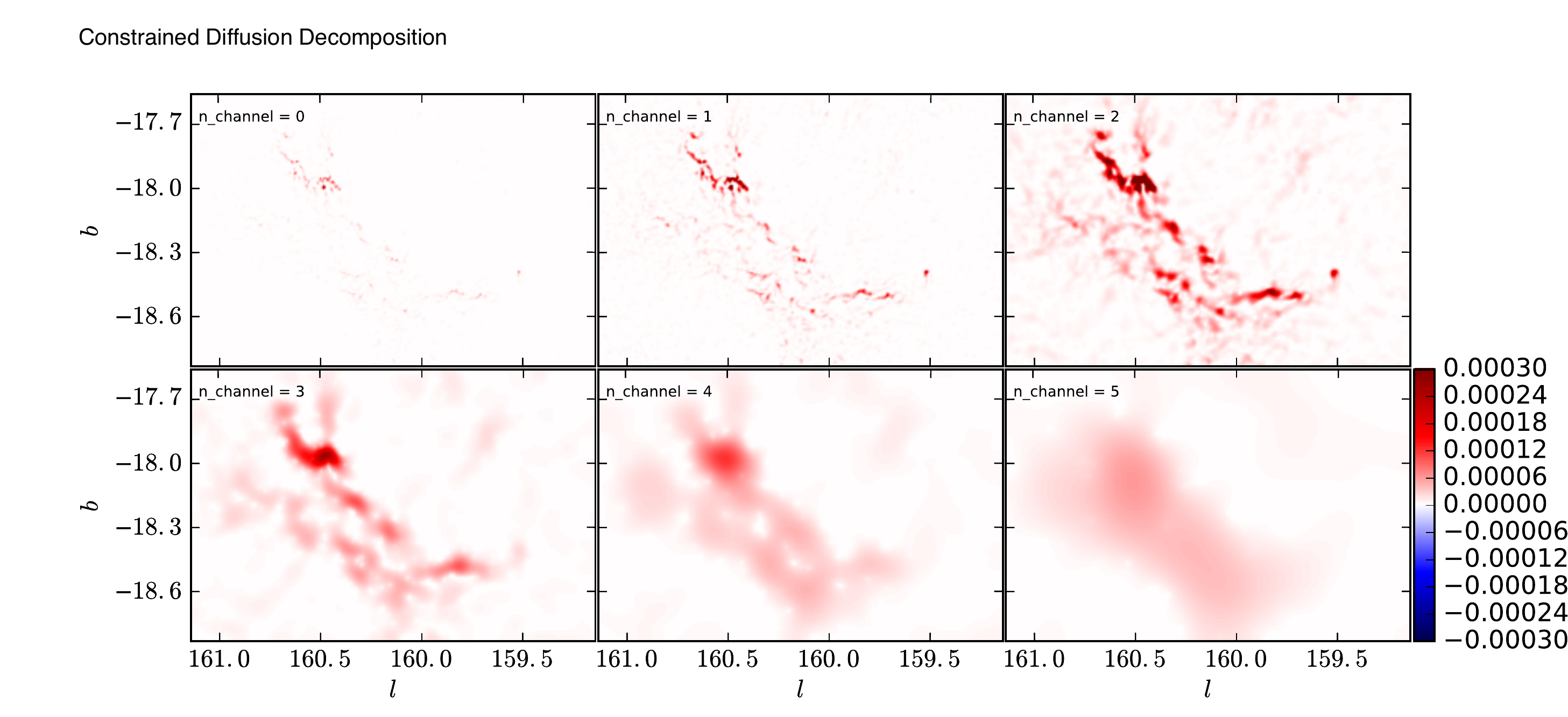}
  \includegraphics[width=0.9 \textwidth]{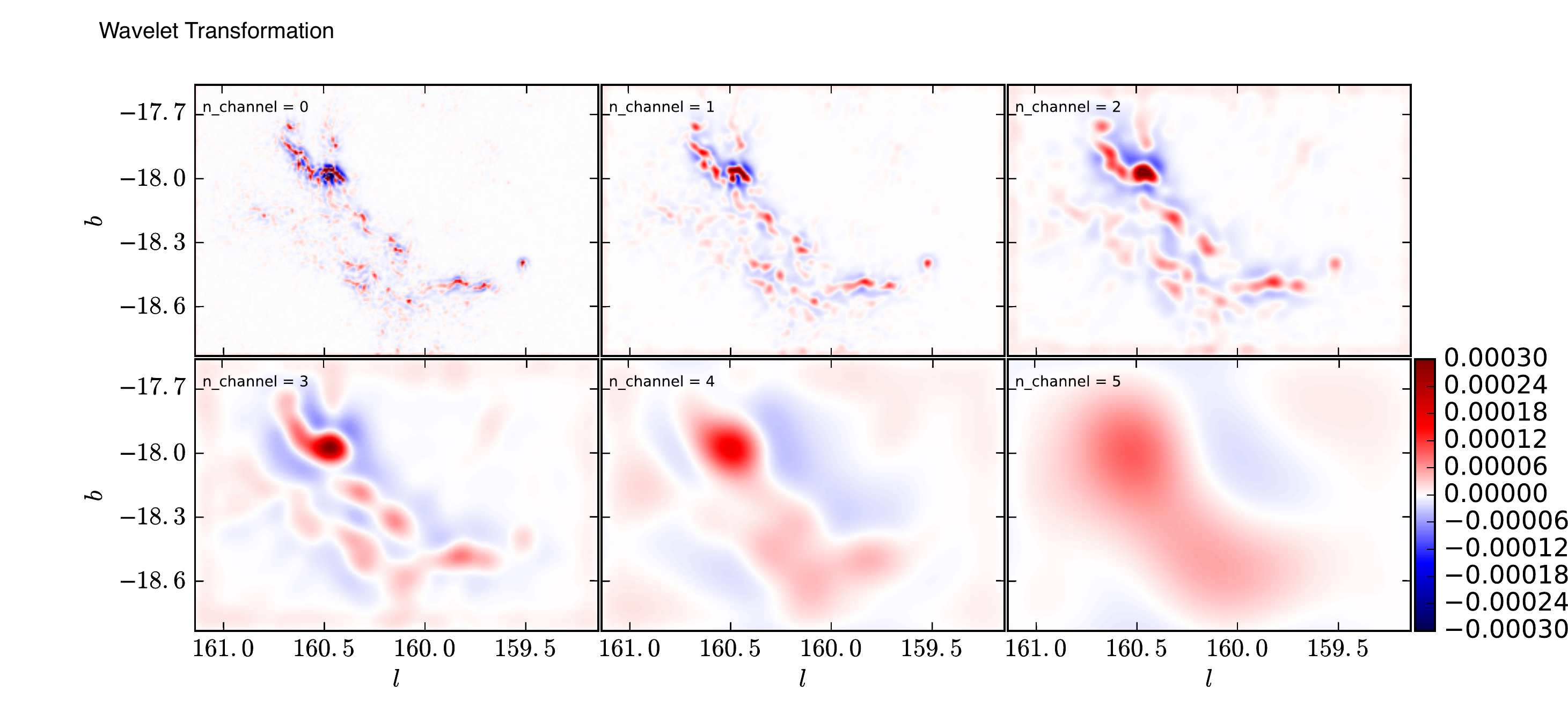}
\caption{Comparison between the results from our constrained diffusion decomposition (upper panels) to those from the wavelet transformation (lower panels). These maps are derived by applying these methods to a surface density map of the IC348 region obtained from \citet{2016A&A...587A.106Z}. \label{fig:ic348} }
\end{figure}

\begin{figure}
  \includegraphics[width=0.9 \textwidth]{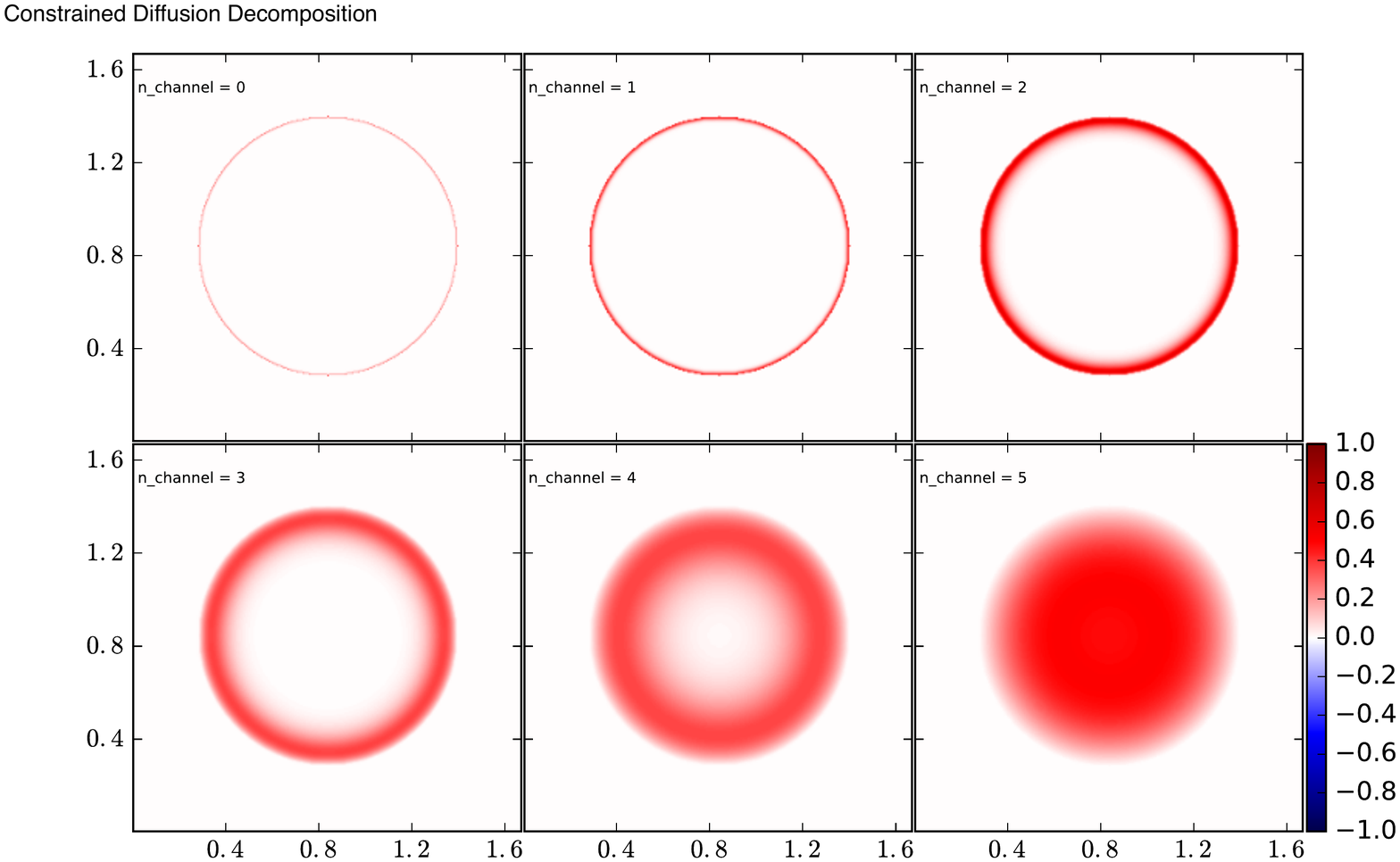}
  \includegraphics[width=0.9 \textwidth]{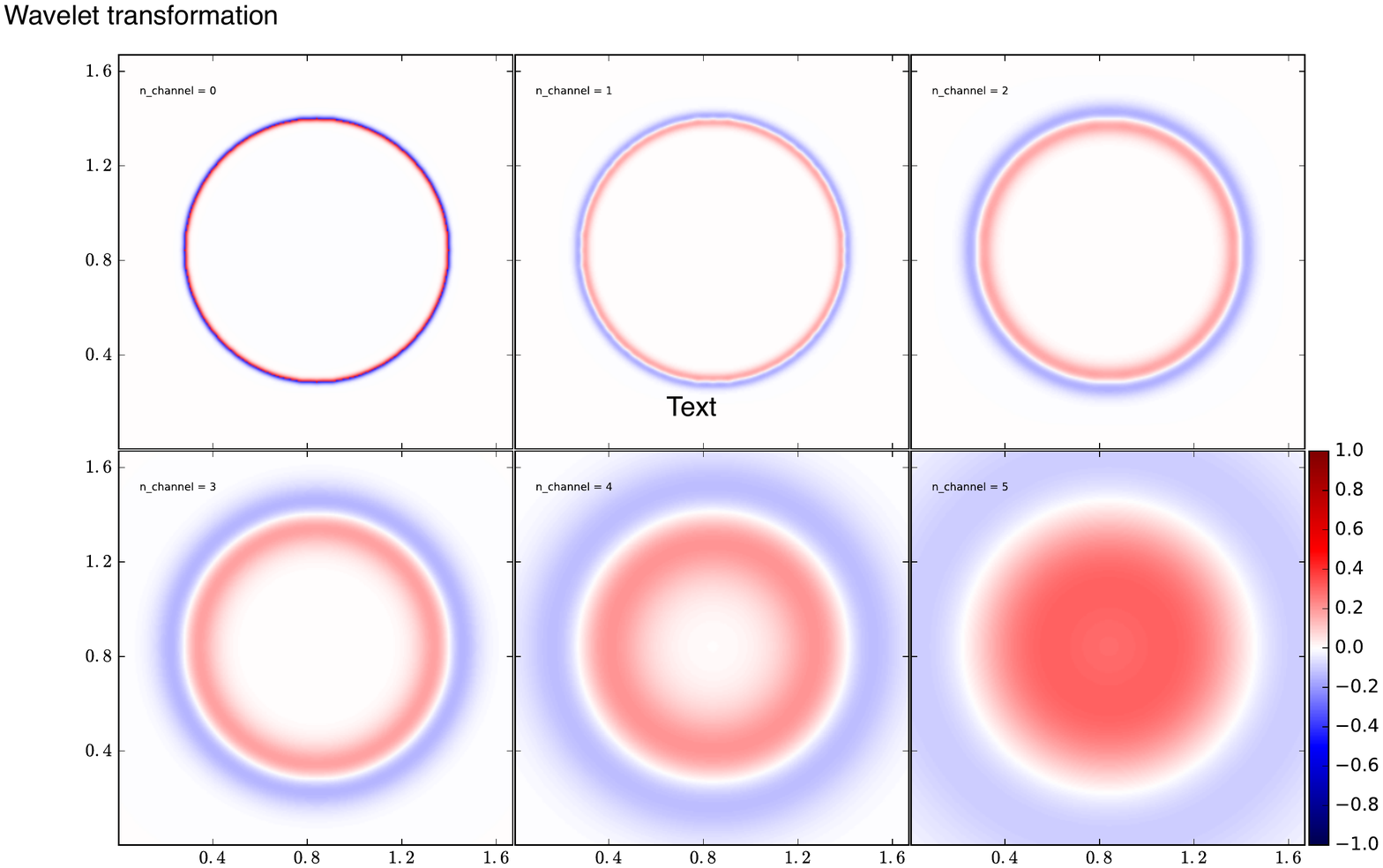}
\caption{Comparison between the results from our constrained diffusion decomposition (upper panels) to those from the wavelet transformation (lower panels). There results are derived by  applying these methods to a map containing a disk of a constant surface density. \label{fig:disk} }
\end{figure}

In Figs. \ref{fig:ic348} and \ref{fig:disk}, we apply our Constrained Diffusion Decomposition to a few examples, and compared the results with those from the wavelet decompositions.

\bibliography{paper}

\end{document}